\documentclass[journal=cgdefu,manuscript=article,layout=twocolumn]{achemso}

\usepackage[version=3]{mhchem} 
\usepackage{graphicx}
\usepackage{float}
\usepackage{latexsym}
\usepackage{amssymb}

\usepackage[section]{placeins}

\newcommand*{\citen}[1]{%
  \begingroup
    \romannumeral-`\x 
    \setcitestyle{numbers}%
    \cite{#1}%
  \endgroup   
}

\author{Federico Giberti}
\affiliation[Department of Chemistry and Applied Biosciences, ETH Zurich]
{Department of Chemistry and Applied Biosciences, ETH Zurich}   
\author{Matteo Salvalaglio}
\affiliation[Department of Chemical Engineering, UCL]
{Department of Chemical Engineering, University College London, Torrington Place, London WC1E 7JE, United Kingdom}
\email{m.salvalaglio@ucl.ac.uk}
\author{Marco Mazzotti}
\affiliation[Institute for Process Engineering, ETH Zurich]
{Institute of Process Engineering, ETH Zurich, CH-8092 Zurich, Switzerland}
\author{Michele Parrinello}
\affiliation[Department of Chemistry and Applied Biosciences, ETH Zurich]
{Department of Chemistry and Applied Biosciences, ETH Zurich}
\alsoaffiliation[Istituto di Scienze Computazionali, Universit\`{a} della Svizzera Italiana]
{Facolt\`{a} di Informatica, Istituto di Scienze Computazionali, Universit\`{a} della Svizzera 
Italiana Via G. Buffi 13, 6900 Lugano Switzerland}

\title{3BrY nucleation: from dimers to \emph{needle-like} clusters}

\begin{document}

\maketitle

\begin{abstract}
The nucleation of 1,3,5-Tris(4-bromophenyl)-benzene (3BrY) occurs through the formation of amorphous intermediates. However, a detailed experimental characterization of the structural and dynamic features of such early-stage precursors is prevented by their transient nature. In this work, we apply molecular modeling to investigate both nucleation and growth of 3BrY clusters from solution. Consistently with experimental findings, we observe that disordered clusters consisting of 10-15 monomers can spontaneously emerge from solution. Such clusters are poorly ordered and can easily fluctuate in size and shape.
When clusters grow to a larger size (up to ~200 monomer units) they develop a markedly elongated morphology, reminiscent of the needle-like 3BrY crystals observed experimentally. 
The growth process is characterized by a continuous rearrangement of ordered and disordered domains accompanied by a dynamical exchange of molecules and oligomers with the solution, in a mechanism resembling the self-assembly of non-covalent supra-molecular fibers. 
\end{abstract}

\section{Introduction}

Nucleation is an elusive and yet pivotal process in crystallization. Shedding light on the molecular basis of nucleation is of paramount importance for improving our current overall understanding of crystallization. The mechanistic picture of nucleation, traditionally portrayed by Classical Nucleation Theory (CNT), has been increasingly challenged as experimental evidences emerged and more articulated descriptions of the process have been proposed \cite{sosso2016crystal}. In this context the so-called \emph{two-step nucleation mechanism}~\cite{vekilov2005two,harano2012heterogeneous,vekilov2010two,kuznetsov1998atomic} is particularly relevant. 
According to the \emph{two-step mechanism}, nucleation begins with the formation of an amorphous precursor in which an embryo of the crystal phase is eventually nucleated. Initial evidences in support of this mechanism were obtained for colloids and proteins nucleating from  solution~\cite{vekilov2005two,vekilov2010two,kuznetsov1998atomic}, but recent studies have suggested that also small organic molecules may follow a similar pathway \cite{salvalaglio2015molecular,salvalaglio2015urea}. 

Harano \emph{et al.} demonstrated that 1,3,5-Tris(4-bromophenyl)-benzene (3BrY) nucleates following a typical \emph{two-step} mechanism. The liquid-like precursors~\cite{harano2012heterogeneous} contain approximatively 15-20 monomers. Nevertheless, mechanistic details of intermediates formation in supersaturated solutions and of their transformation in nuclei of the crystal phase remain unclear as they elude typical length and time scales that can be simultaneously probed in experiments \cite{vekilov2012crystal}. 
Due to their intrinsic atomic resolution, Molecular Dynamics based techniques can be applied to study nucleation with the twofold aim of developing a molecular-level understanding of mechanisms and of helping in the explanation and rationalization of experiments. 

Unfortunately, MD simulations are limited by the size of the discrete time step needed to integrate Newton equations of motion, usually 1-2 fs. With present-day computational resources, this limits the accessible timescale in the $\mu$s for most processes. 
Since nucleation is an activated process, the probability of observing the formation of a nucleus on this timescale is usually negligible, unless drastic conditions are imposed such as deep undercooling or very high supersaturation\cite{ten1996numerical,trudu2006freezing}. Imposing such extreme conditions can substantially alter the nucleation process and induce profound effects on the nucleation mechanism\cite{trudu2006freezing}. 

The sampling of activated processes, associated with long timescale events is a ubiquitous problem in MD and more generally in computational physical chemistry. A variety of  enhanced sampling techniques have been proposed to overcome such 
limitations\cite{huber1994local,torrie1977nonphysical,laio2002escaping,dellago2002transition}. 
In this work, we apply Well-Tempered Metadynamics (WTmetaD) \cite{Barducci2008} to enhance the sampling of the reversible formation of a 3BrY nucleus. Metadynamics is an adaptive biasing technique in which visiting already sampled conformations is discouraged via the introduction of an external time-dependent repulsive potential. 
The application of such a potential allows to achieve an extensive sampling of the configurational space and allows recovering a Free Energy Surface (FES)\cite{barducci2011metadynamics,giberti2015metadynamics}. 

In this work using both WTmetaD and MD simulations, we provide a detailed description of the formation of 3BrY clusters in solution, which lead to the formation of a crystal nucleus. We begin by analyzing 3BrY dimerization, proceeding then with the investigation of 3BrY clusters formation and finally studying the growth of a supercritical 3BrY nucleus. It should be noted that 3BrY has a solubility in Ethanol which is safe to consider below the 1$\times$10$^{-3}$ mol l$^{-1}$. This restricts our computational analysis in a regime of high supersaturation, which nevertheless corresponds to experimentally relevant conditions. The paper is divided into three parts: at first the computational methods used are introduced, then the results are reported, and finally, conclusions are drawn.

\section{Methods}
\label{sec:method3br}

\paragraph{Simulations summary}
Both unbiased MD and WTmetaD simulations were carried out to investigate the formation of clusters of 3BrY in solution. 
A summary of the simulations performed is reported in table \ref{tab:3brysimulations} where the sampling method is indicated together with the total production time and the solvent used.
Simulations A, B, and C were equilibrated in the NPT ensemble with the Parrinello-Rahman barostat\cite{Parrinello}. After equilibration, production runs were carried out in the NVT ensemble, where the temperature was controlled with the Bussi-Donadio-Parrinello thermostat \cite{Bussi}. Long-range electrostatics were computed with the particle-mesh Ewald (PME) method, and the time-step was set to 2 fs using the LINCS algorithm (simulations A, B, and C) to constrain bond lengths.
These simulations were performed with GROMACS-4.5.5 \cite{gromacs1,gromacs2,gromacs3,gromacs4}, patched with PLUMED-1.3\cite{PLUMED}. Simulation D was conducted in the NVT ensemble on a GPU workstation, using \emph{pmemd}, the GPU-optimized engine from the AMBER suite of programs\cite{gtz2012routine,salomon2013routine}. Also, in this case, long-range interactions were computed with the Particle-Mesh Ewald approach, while bond lengths were constrained using the SHAKE algorithm thus allowing to run simulations with a time-step of 2 fs. In all simulations, the cutoff for short range interactions was set to 1.0 nm. 

\begin{table*}
\centering
\begin{small}
\caption{Details for the 3BrY simulations. For the metadynamics simulations, the $\gamma$ factor was chosen so to quickly temper the bias on the order of tens of ns and sample the whole CV space. The width of the Gaussian was taken to be smaller than the thermal fluctuation of the CV in the minima, to be able to resolve them correctly.}
\label{tab:3brysimulations}
\begin{tabular}{ccccccc}
label& N  3BrY    & solvent&  time   & Method   & $\gamma$ factor & width of \\ 
     & monomers   &        & [$ns$]  &          &             &Gaussian \\   \hline
A    & 2          & EtOH    &    50   &  WTmetaD &     5       &  0.05\\
B    & 32         & EtOH    &   250   &  WTmetaD &     20      &   1\\
C    & 32         & EtOH    &   50    &  MD      &      -      &   \\
D    & 380        & MetOH   &  100    &  MD      &      -      &    \\
\hline
\end{tabular}
\end{small}
\end{table*}

\paragraph{3BrY forcefield} To model 3BrY molecules in explicit Methanol and Ethanol solutions, we have used the Generalized Amber Force Field (GAFF) \cite{GAFF,GAFF1}. Partial atomic charges were computed using the restrained electrostatic potential formalism \cite{RESP}, evaluated at the DFT level with B3LYP/6-31G(d,p) basis functions. DFT calculations were performed with GAUSSIAN09 \cite{g09}. A comparison of the crystal structure obtained with this forcefield and the experimental structure obtained via X-ray diffraction at 136 K was reported in a previous work\cite{salvalaglio20141,beltran20021}.

\paragraph{Enhanced Sampling with metadynamics}To extensively sample the formation and disruption of clusters in solution, we  applied WTmetaD \cite{Barducci2008}. 
In plain metadynamics\cite{laio2002escaping}, an external bias potential $V_M(S,t)$ acting on a set of Collective Variables (CV) $S$ is iteratively constructed during the simulation, in such a way that it discourages the sampling of regions of the $S$ space that have already been visited. An estimate of the FES along $S$ is also obtained as the negative of the bias deposited during the MD simulations. 
In WTmetaD, a smooth convergence of the FES is promoted by decreasing the amount of biasing potential introduced in the system during the simulation. The coefficient used to converge the bias ($\gamma$ factor) is reported in table \ref{tab:3brysimulations}. The interested reader can find an exhaustive description of the method in Refs. \citen{barducci2011metadynamics},\citen{valsson2016enhancing} and in references therein; a brief report on its applications to crystal nucleation can be found in Ref. 
\citen{giberti2015metadynamics}.

\paragraph{Collective Variables} In order to enhance the formation of ordered nuclei, we have constructed the WTmetaD bias as a function of the order parameter $S$ that was developed to enhance the sampling of nucleation events in molecular systems ~\cite{giberti2015insight}. 
The variable $S$ is expressed as the sum of single particle contributions $\Gamma_i$: 
\begin{equation}
S = \sum_i^N \Gamma_i  = \sum_i \rho_i \theta_i
\end{equation}
Where $\rho_i$ accounts for the local density and $\theta_i$ accounts for the orientation of solute molecules. 
The first term, $\rho_i$ is expressed as: 
\begin{eqnarray}
\rho_i&=&\frac{1}{(1+e^{-b(n_{i}-n_{cut})})}   
\label{eq:1}
\end{eqnarray}
where $n_i$ is the coordination number and $n_{cut}$ a reference threshold; $n_i$ is given by: 
\begin{equation}
n_i = \sum_j^N f_{ij}= \sum_j^n\frac{1}{(1+e^{a(r_{ij}-r_{cut})})}
\label{eq:2}
\end{equation}

The second term $\theta_i$ quantifies if molecule $i$ and its neighbors possess a relative orientation $\vartheta_k$ compatible with one of those observed in the crystal. 
This is measured via the angle $\vartheta_{ij}$, formed between two intramolecular vectors placed on each molecule. 
While in a perfect crystal at 0 K the angle $\vartheta_{ij}$ is equal to the angle $\vartheta_k$, at finite temperature we observe a distribution of angles, which is peaked at $\vartheta_k$ with a finite variance. The angular part is thus expressed as a sum of  $k_{max}$ Gaussian functions equal to the number of possible orientations, each one peaked at a different $\vartheta_k$ with variance $\sigma_k$. To constrain this contribution between 0 and 1, the whole sum is normalized by the number of neighbors $n_i$.

\begin{equation}
\theta_{i}=\frac{1}{n_i} \sum_j^N f_{ij} \sum_k^{k_{max}}
e^{-\frac{(\vartheta_{ij}-\bar{\vartheta}_k)^2}{2\sigma_k^2}}
\end{equation}

\begin{figure}[ht!]
\centering
\includegraphics[width=0.35\textwidth]{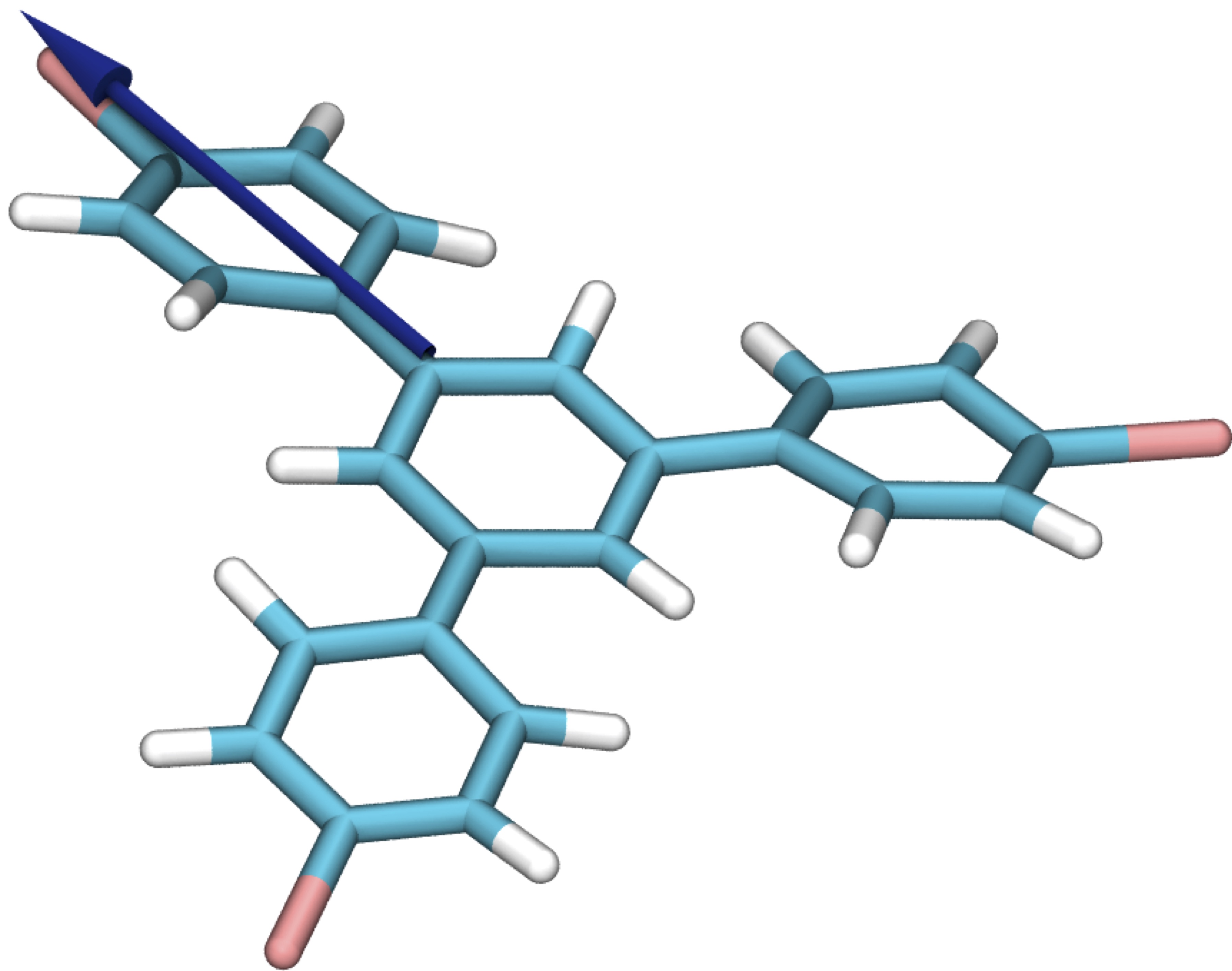}
\caption{Graphical representation of the intramolecular vector used to define the relative orientation of adjacent 3BrY molecules.}
\label{fig:3brymolecule}  
\end{figure}

The resulting $\Gamma_i$ approaches 1 for a molecule in a crystal-like environment, while it reduces to 0 otherwise. Taking advantage of this formulation the variable $S$ also approximates the number of crystal-like molecules in the system. 
Knowledge of the crystalline structure is key to define the parameters of $S$. To our knowledge, 3BrY does not display polymorphism and $S$ was defined considering the molecular environment of 3BrY in its only known crystal structure\cite{salvalaglio20141}. 
The parameter $r_{cut}$ of the switching function was chosen in such a way to be placed in the first minimum of the intermolecular $\overline{CC}$ radial distribution function. Since 3BrY molecules exhibit a D$_{3h}$ symmetry, three angles were defined to identify the crystal-like configuration, with a width of the Gaussian functions of $27^{o}$. 
A summary of the parameters is reported in Tab. \ref{tab:3bryvariable} while a representation of the intramolecular vector defining the orientation of a single molecule is illustrated in Fig. \ref{fig:3brymolecule}.

\begin{table}[ht!]
\centering
 \caption{Parameters used for the calculation of $S$.}
\begin{tabular}{ccccccc}
\label{tab:3bryvariable}
label   & $r_{cut}$& $n_{cut}$  & $\overline{\vartheta}_k$  & a& b \\
&  \small [nm] &  &  \small [deg] & \small [nm$^{-1}$] &  & \\ \hline
A   & 2.2  &   1    &  0; 120; 240        & 20 & 20  \\ 
B,C,D   & 0.6  &   1    &  0; 120; 240        & 20 & 20  
\end{tabular}
\end{table}

This variable can identify molecules embedded in a crystalline environment, however, it is unable to discriminate between a cluster of disordered molecules and a homogeneous solution, as both states are characterized by values of $S$ approaching zero. To differentiate between them a second variable, $Z$, was introduced, which is the sum of the density terms $\rho_i$: 
\begin{equation}
 Z= \sum_i \rho_i 
\label{eqn:SMCC}
\end{equation}
This CV is proportional to the number of molecules possessing a coordination number greater than $n_{cut}$ and hence can discriminate between the presence and the absence of clusters in the system. This variable was used only for analysis purposes and not to introduce an external bias in MD simulations.

\section{Results}
\subsection{3BrY dimerization}

To improve our understanding of the dominant interactions leading to the formation of amorphous 3BrY precursors, we have investigated the self-association of two 3BrY molecules in solution in a box of 325 ethanol molecules using WTmetaD (simulation A, Tab.~\ref{tab:3brysimulations}).

\begin{figure*}[ht!]
\centering
\includegraphics[width=0.75\textwidth]{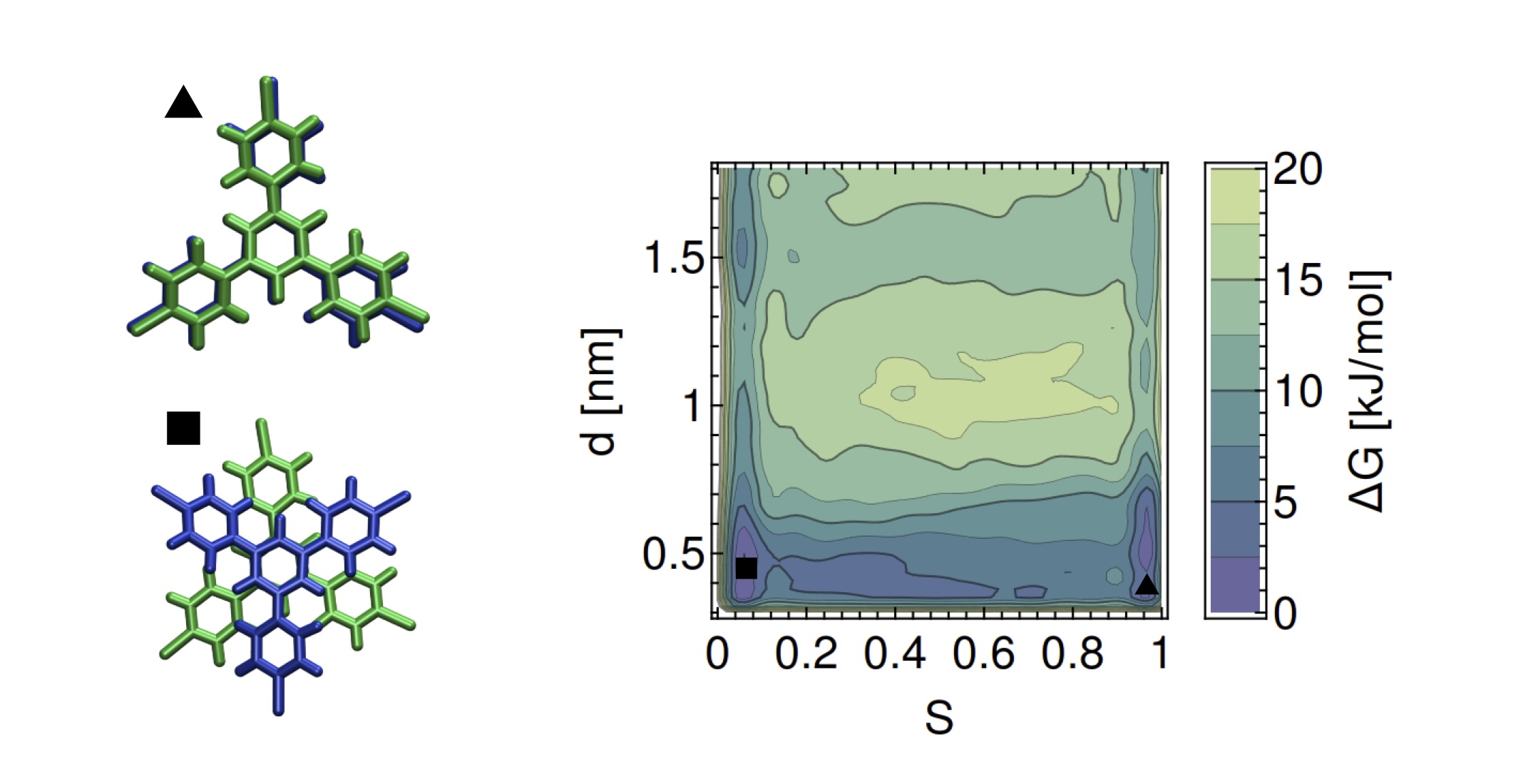}
\caption{Simulation A. \emph{left}) Two limiting dimer configurations: eclipsed ($\blacktriangle$) and staggered ($\blacksquare$) B) FES associated with the formation of a 3BrY dimer in ethanol solution. The FES illustrates how the dimerization process is favored and the interconversion between different dimer configurations is characterized by a small barrier. This information suggests that 3BrY dimers can reorganize and reorient with small energetic costs, hence on short timescales.}
\label{fig:3brydimer}
\end{figure*}

The formation and disruption of a dimer were enhanced through WTmetaD. The bias potential was applied to two CVs, namely $d$ the distance between the centers of mass of two 3BrY molecules, and $S$ the variable defined in the previous section.
The FES in Fig.\ref{fig:3brydimer} clearly shows that the formation of dimers is thermodynamically favored, and the main driving force to dimerization can be identified in the establishment of interactions between phenyl rings.
When bound, the two 3BrY molecules can rotate relative to one another. In doing so the conformations in which the two molecules are arranged either in an eclipsed (S$\simeq${1}) or a staggered (S$\simeq${0}) configuration are slightly favored energetically. The FES reported in Fig.\ref{fig:3brydimer} shows that such rotation is associated to free energy barriers of the order of 2 $k_BT$, and therefore can take place almost unhindered. The eclipsed and staggered limiting structures are reported in Fig. \ref{fig:3brydimer}. 
A similar behavior was observed in our previous study of 3BrY dimerization in methanol. In that case, even if the dimer configuration was found to be more stable than in ethanol, similar configurational freedom was reported \cite{salvalaglio20141}.

\subsection{Structure, size and shape fluctuations of disordered 3BrY precursors}

\paragraph{Size and shape of precursors.}To investigate the spontaneous formation of 3BrY clusters, an unbiased simulation of 32 3BrY molecules was performed in a box of about 3000 ethanol molecules at 273K. (Simulation C, Tab.\ref{tab:3brysimulations}). 

After an initial equilibration, molecules aggregate in some small clusters, exhibiting remarkable variety in both size and shape. Clusters were identified applying a Depth First Search algorithm, as discussed in detail in Ref. \citen{tribello2017analyzing}. 
In order to describe their structure, we define two clustering criteria for 3BrY molecules, i.e. first according to their Cartesian distance, hence neglecting their relative orientations, and secondly depending on their distance and their degree of local order, measured as the value of the mono-molecular order parameter $\Gamma_i$ defined in the previous section. 
In the first case pairs of molecules were considered as belonging to the same cluster if they were closer than a 0.95 nm. In the second case in addition to this criterion molecules were required to possess a $\Gamma_i$ value larger than  0.3. Such value was chosen to include in the clusters also molecules that are at the surface of ordered clusters and not only at their core, where $\Gamma_i\approx 1$. 
From our previous work \cite{salvalaglio20141}, we are aware that 3BrY molecules can arrange in multi-columnar packings where 3BrY molecules belonging to different columns lie approximately in the same plane. Our choice of clustering parameters is such that multi-columnar configurations can be identified. Although relatively rare in small clusters,  these arrangements can be found in larger assemblies, in which several columns of stacked 3BrY molecules interact with each other.
These two clustering procedures yield complementary information, namely on the size distribution of the clusters, and on their level of order, respectively. 

\begin{figure*}[ht!]
\centering
\includegraphics[width=0.75\textwidth]{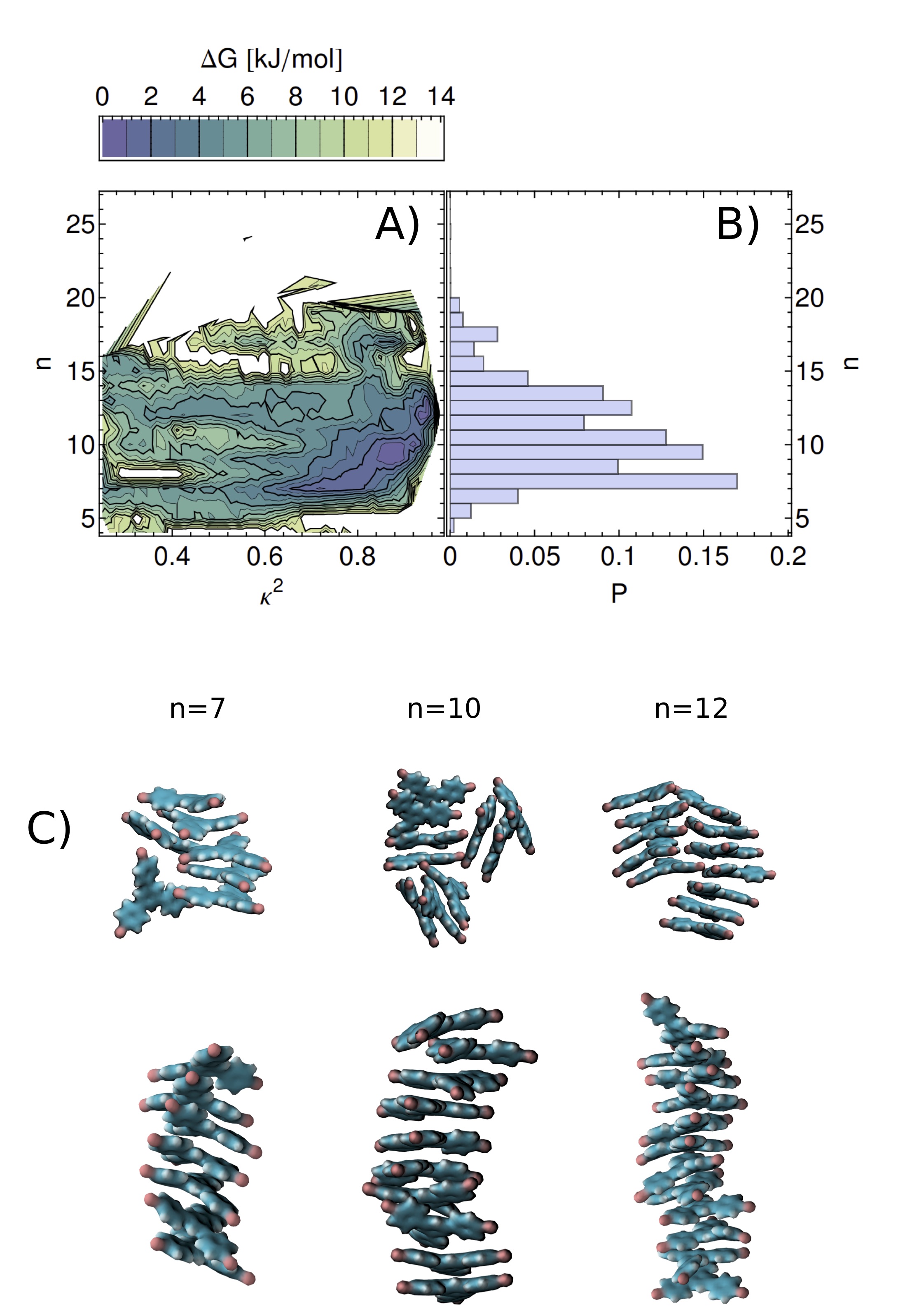}
\caption{Simulation B. A) FES obtained from the unbiased simulations as a function of the shape anisotropy $k^2$ and the number of 3BrY molecules in a cluster $n$. B) The probability distribution of the size of the largest cluster in the unbiased simulations obtained as a function of the number of monomers $n$. C) cluster structures illustrating the shape flexibility possessed by the precursor: on the top row compact isotropic structures characterized by a $k^2<0.5$ are reported, while on the bottom row elongated, rod-like structures characterized by $k^2>0.8$ are depicted.}
\label{fig:3brylargest}
\end{figure*}

As expected, the formation and disruption of 3BrY clusters in solution occurs frequently and without the need for large energy fluctuations.
The probability distribution of the size of the largest cluster is reported in Fig.\ref{fig:3brylargest}b. The distribution ranges from 5 to 20 molecules and is centered at 10 monomers. Such size range is consistent with the size of the precursors experimentally identified by Harano et al. \cite{harano2012heterogeneous}. We observe that clusters possess a fluxional nature, and they tend to spontaneously fluctuate both in shape and size.

In order to quantitatively characterize the cluster shape we compute the anisotropy index $k^2$, function of the eigenvalues of the inertia tensor $\lambda_x, \lambda_y,$ and $\lambda_z$:   
\begin{equation}
 k^2 = 
\frac{3}{2}\frac{\lambda_x^4+\lambda_y^4+\lambda_z^4}{
(\lambda_x^2+\lambda_y^2+\lambda_z^2)^2 } -\frac{1}{2}
\label{eqn:shape}
\end{equation}

The anisotropy index $k^2$ approaches 0 for a sphere and 1 for a straight line\cite{theodorou1985shape}.
The FES computed as a function of $k^2$ and $n$, reported in Fig. \ref{fig:3brylargest}a, provides a quantitative measure of structural fluctuations of 3BrY clusters both in size and shape. For instance, the FES reported in  Fig. \ref{fig:3brylargest}a shows a minimum in the region where $8<n<15$ and values of $k^2$ approaching 1. 
This indicates that within the most probable cluster size (~10 3BrY molecules) elongated cluster structures are thermodynamically favored over compact isotropic configurations.  
To better illustrate structural differences between elongated and compact structures, typical cluster morphologies with size ranging from $n=7$ to $n=12$ are reported in Fig. \ref{fig:3brylargest}c. In the top row of Fig. \ref{fig:3brylargest}c compact cluster configurations ($k^2<0.5$) are reported, while in the bottom row elongated columnar clusters ($k^2>0.8$) are shown. Our analysis, in agreement with the experimental findings of Harano et al. \cite{harano2012heterogeneous}, indicates that 3BrY crystal nucleation is likely to proceed through the formation of intermediate clusters. Furthermore, our findings suggest that the wide population of clusters fluctuating in both size and shape is dominated by elongated clusters characterized by a columnar arrangement of 3BrY molecules. 

\paragraph{Structure of precursors.} Though fluctional and rather amorphous, cluster structures sampled in MD simulations are not completely disordered.
Locally ordered domains constituted of 2 to 4 adjacent molecules are in fact consistently observed in our simulations.
To quantify the role and presence of such ordered domains in 3BrY clusters, we compute two additional descriptors of the cluster structure: the fraction of ordered 3BrY molecules in each cluster $\eta$, and the number of ordered fragments in each cluster $n_{OF}$.
 
\begin{figure*}[ht!]
\centering
\includegraphics[width=0.75\textwidth]{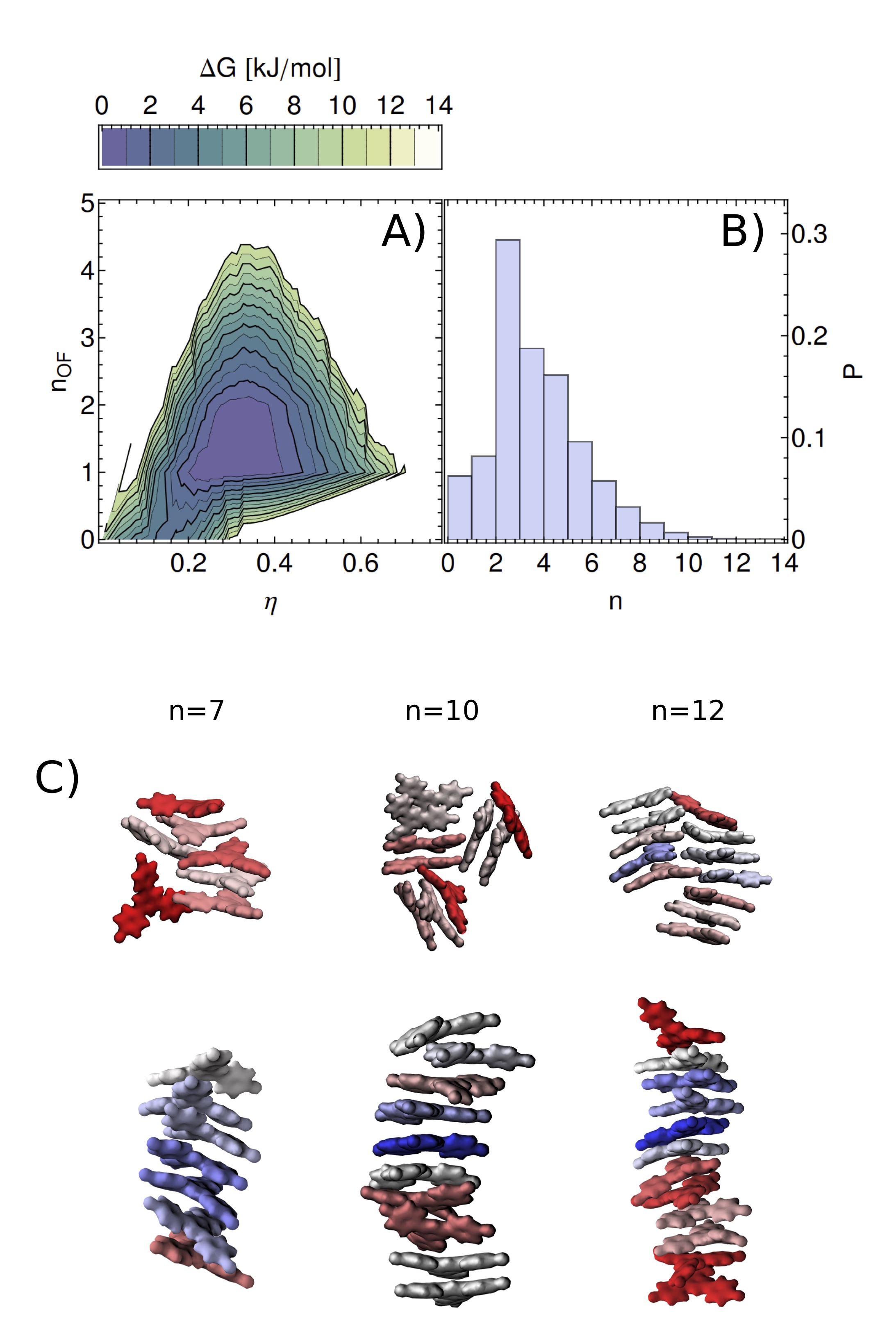}
\caption{Simulation B. A) FES as a function of the ratio of ordered molecules $\eta$ and the number of ordered fragments $n_{OF}$ in the precursor. 
The stable minimum corresponds to structures with $\eta\approx$ 0.35. B) Size distribution of an ordered sub-domain in the largest cluster. C) Examples of structures of different sizes are reported, in which each molecule is colored according to its $\Gamma_i$ parameter (a red-white-blue color scale was used with $\Gamma_i=$0 corresponding to red and $\Gamma_i=$1 corresponding to blue).}
\label{fig:3bryorder-disorder32}
\end{figure*}

The FES as a function of $n_{OF}$ and $\eta$ shows a single basin. Its global minimum, representing the most likely configuration, corresponds to aggregates characterized by a single crystal-like domain, possessing an ordered fraction $\eta$=0.30. Nevertheless, fluctuations in the degree of internal order $\eta$ are associated with relatively small free energy fluctuations. 
Consistently with the results obtained for the 3BrY dimers, such structural fluctuations are typically associated with rotations of 3BrY molecules around their $CV_3$ symmetry axis. 
In Fig.\ref{fig:3bryorder-disorder32}b we report the probability associated with the number of ordered molecules within each cluster. Such histogram exhibits a maximum in the region between 2 and 3 molecules. 
In Fig. \ref{fig:3bryorder-disorder32}c the same cluster structures reported in Fig. \ref{fig:3brylargest}c are shown. For each structure, the molecules are colored according to the value of their $\Gamma_i$ parameter. In this figure, it can be readily seen that columnar clusters are more ordered and possess a single ordered domain. This analysis on the structure of the clusters allows concluding that, though maintaining a fluxional and overall disordered character, clusters of 3BrY molecules are likely to possess at least one domain of molecules arranged in a crystal-like oriented configuration.  Such domains, embedded in clusters spontaneously forming from solution represent the smallest structural unit displaying a crystal-like character. 

\begin{figure*}[ht!]
\centering
\includegraphics[scale=0.10]{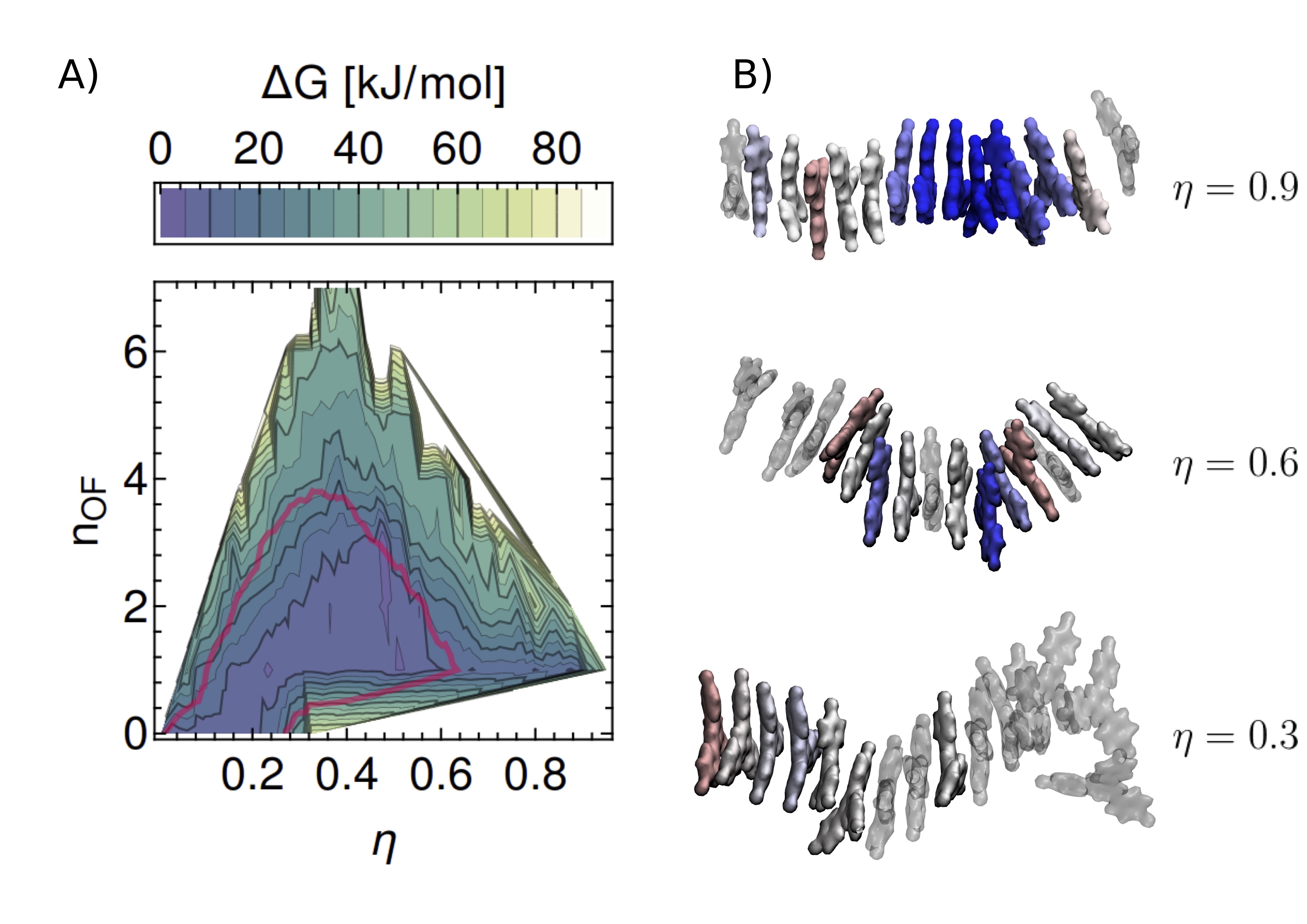}
\caption{Simulation C. A) FES reweighed from the WTmetaD simulation reported as a function of $\eta$ and $n_{OF}$. The red contour is reported to illustrate the limit of the unbiased MD simulation. While the global minimum is still at $\eta$ about 0.35, clusters with an ordered fraction greater than 0.6 can now be obtained. B) Examples of clusters obtained from the WTmetaD simulation were molecules with $\Gamma_i>$ 0.3 were colored according to their $\Gamma_i$  
(a red-white-blue scale was used where $\Gamma_i$=0.3 corresponds to red and $\Gamma_i$=1 corresponds to blue). Molecules with a $\Gamma_i<$0.3 were reported in transparent gray.}
\label{fig:smac-smcc}
\end{figure*} 

\paragraph{From local to long-range ordered cluster with WTmetaD}

While allowing for a significant sampling of the cluster population in explicit ethanol solution, MD simulations do not allow to sample the transition towards structures characterized by a long range crystal-like arrangement. 

This is due to the activated nature of the nucleation process as well as to finite size effects typical of MD-based simulations in which the chemical potential of the mother phase cannot be considered constant during the nucleation process\cite{salvalaglio2015molecular}. Both these limitations can be effectively overcome using an enhanced sampling method such as WTmetaD\cite{salvalaglio2015molecular,salvalaglio2015urea,salvalaglio2016overcoming}. 

To this aim we have carried out a WTmetaD simulation of the nucleation process (simulation B), utilizing as a CV the function $S$ described in detail in the methods section.  The width of repulsive Gaussian functions and the $\gamma$ factor are reported in Tab. \ref{tab:3brysimulations}.
Reweighing \cite{bonomi2009reconstructing} the statistics gathered from WTmetaD simulations, the FES in the space of $n_{OF}$ and $\eta$ was computed. 

The FES, obtained from WTmetaD and shown in Fig. \ref{fig:smac-smcc} is consistent with that obtained from unbiased MD simulations and confirms that in ethanol solution 3BrY molecules can easily cluster. The size, shape and structural properties of clusters in this basin are consistent with those obtained from simulation C. 

WTmetaD, however, allows enhancing both structural and size fluctuations on a larger energy scale, allowing sampling long-range ordered structures comprising of up to 30 3BrY molecules in a single crystal-like domain stabilized by periodic boundary conditions.  

The barrier separating such structures from the basin representing the solution populated by smaller clusters is  $\approx$ 15 kJ/mol. Such an estimate represents a lower bound for the real barrier as in our finite-sized system the largest ordered clusters are stabilized by periodic boundary conditions.
Nonetheless, it emerges that a critical step in the nucleation of a crystal-like structure from disordered clusters is hampered by the formation of a \emph{single domain} of ordered molecules. 

This finding suggests that the two-step nucleation process, rather than implying a sudden and complete conversion of the disordered precursor into a crystal-like structure, proceeds through the growth of crystal-like domains embedded in larger 3BrY clusters that still maintain a large number of disordered molecules.
To further verify this finding, the growth of a larger cluster from supersaturated solution were simulated. Results are discussed in the following section.

\begin{figure*}[ht!]
\centering
\includegraphics[width=0.75\textwidth]{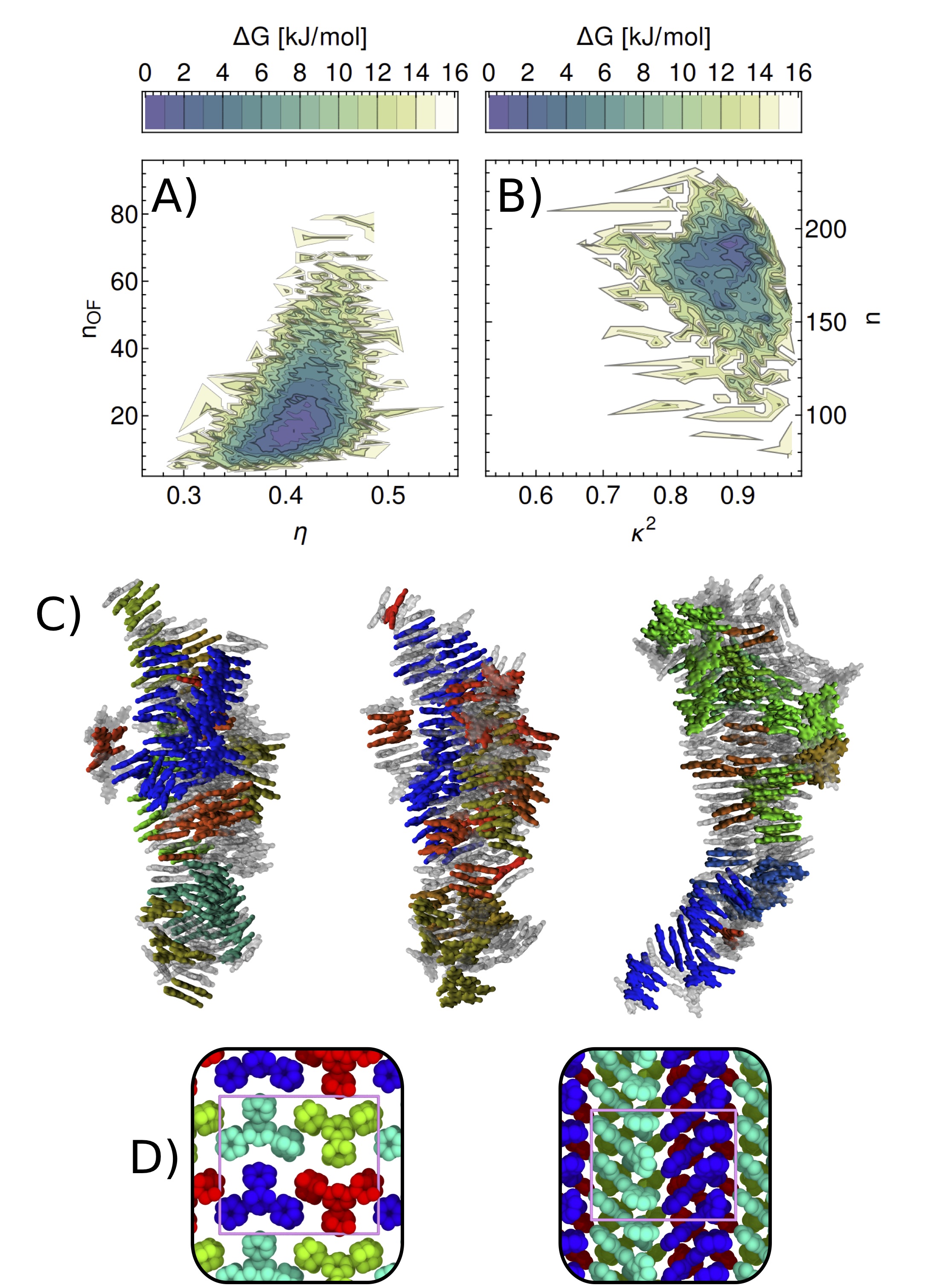}
\caption{Simulation D. A) FES as a function of the order ratio in the cluster $\eta$ and the number of ordered fragments $n_{OF}$. As expected the order in the precursor increases with its size. B) FES as a function of the shape anisotropy $k^2$ and the number of 3BrY molecules in the largest cluster $n$. While the structure is characterized by significant disorder and the cluster has a markedly fluctional character, its shape remains needle-like. C) Three snapshots of the largest cluster, where ordered domains are highlighted in color according to their size using a red-green-blue color scale with red corresponding to small ordered fragments and blue to large ones. Molecules not associated with any locally-ordered fragment are reported in transparent gray. In both cases, only the steady state part of the trajectory (t>40 ns) was analyzed to compute the FESs.}
\label{fig:shapeanysotropy}
\end{figure*}  

\subsection{Growth of a supercritical cluster}

At first, a cluster structure of 60 3BrY molecules was obtained from an unbiased MD simulation performed in a highly supersaturated simulation box. 
Then, such a cluster was embedded in a larger box in which 320 3BrY additional molecules were properly solvated in methanol. 
The low solubility of 3BrY in methanol, with respect to ethanol, diminishes the probability of dissolving the initial seed, thus allowing to observe an unbiased trajectory of a large 3BrY cluster growing. For the same reasons methanol was chosen as a solvent for the TEM analysis by Harano \emph{et al.} \cite{harano2012heterogeneous}.

\begin{figure}[ht!]
\centering
\includegraphics[width=0.45\textwidth]{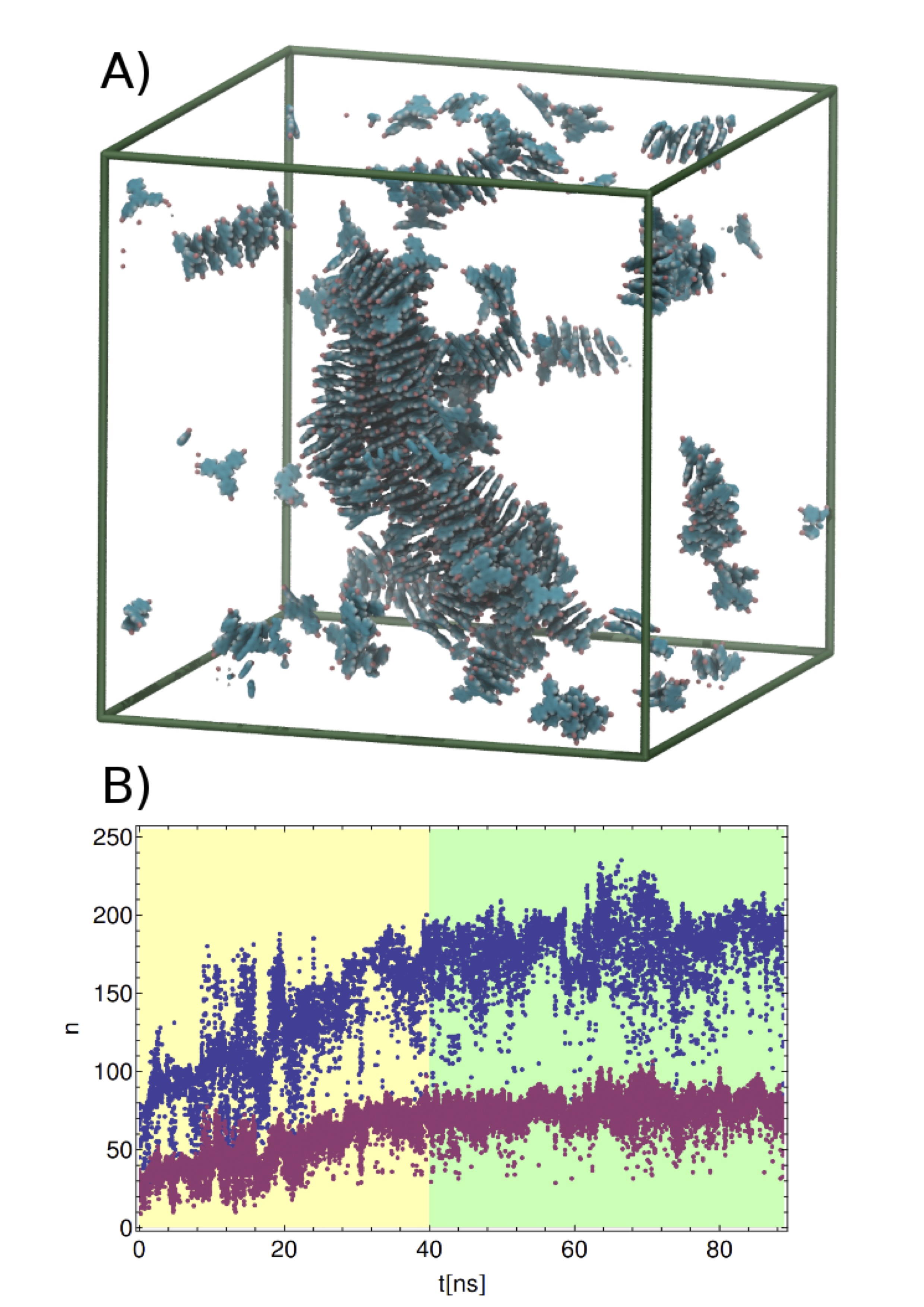}
\caption{Simulation D. A) Equilibrated structure of the largest cluster, with smaller clusters surrounding it in solution. Solvent molecules are not reported for clarity. Panel B). Number of molecules in the largest cluster (blue) and number of ordered molecules (magenta) as a function of time. The growth part of the simulation is highlighted in yellow, while the steady state part, used to calculate the FESs is highlighted in green.}
\label{fig:3brycluster}
\end{figure}

From the starting configuration, the largest nucleus rapidly grows by incorporating additional 3BrY molecules. Simultaneously, while the larger cluster grows, the 3BrY molecules in solution develop a population of clusters different in shape and size. As shown in Fig. \ref{fig:3brycluster}, after approximately 50 ns, the simulation reaches a steady state, in which about 200 molecules of 3BrY form small clusters, whereas the largest cluster, grown from the initial seed, includes about 180 molecules. 

It is now interesting to apply the same analysis carried out for the previous simulations on the largest cluster, to elucidate and quantify its structure, size and shape. 

To this aim we lump the structural information in the FES calculated as a function of the order ratio $\eta$ and the number of ordered fragments $n_{OF}$,  reported in Fig.\ref{fig:shapeanysotropy}a. 
The steady state structure, identified by the global minimum of the FES, is characterized by an order ratio $\eta=0.40$, and is consisting of multiple ordered and disordered domains. The average number of ordered domains within the largest cluster is between 15 and 20, composed by 4 to 20 3BrY molecules each. Examples of the cluster structure at steady state, in which the domains with higher order are highlighted in color, are reported in the lower panel of Fig.\ref{fig:shapeanysotropy}c.  

It is interesting to notice that the growth of the largest cluster takes place through both the addition of single 3BrY molecules and the incorporation of smaller clusters. 
Moreover, even if such cluster is an order of magnitude larger than the typical disordered precursors spontaneously assembled in solution, it lacks a well-defined bulk. 
In fact, most of the molecules are effectively in contact with the solution and therefore should be considered part of the cluster solution interface. 

Due to these characteristics, the structure of the growing cluster is far from static: it is instead continuously rearranging and exchanging molecules with the solution.
Such characteristics remain true even when the cluster reaches a steady state configuration. 

It is interesting to note that the overall shape of the largest 3BrY cluster obtained at steady state in simulation D is reminiscent of the morphology exhibited by the needle-like 3BrY crystals observed in STM experiments \cite{harano2012heterogeneous}.
Such cluster is in fact dominated by domains of 3BrY molecules arranged in columnar packings that tend grow by propagating the apolar stacking characterizing the intermolecular interactions of 3BrY molecules, hence producing an elongated shape. 

Also in this case, in order to quantitatively assess the shape of the largest cluster, the anisotropy index $k^2$ was computed.
The result is illustrated in the FES  reported as a function of the size of the largest cluster $n$ and $k^2$ in Fig. \ref{fig:shapeanysotropy}b.
The global minimum in the FES corresponds to the steady-state configuration of the largest cluster, characterized by a number of 3BrY molecules $n=180$ and a value $k^2=0.9$.

Our analysis shows that particles grown from supercritical clusters possess ordered, crystal-like domains embedded in a disordered environment. The growth process takes place through a dynamical incorporation of monomers and small oligomers. Despite their structural heterogeneity, 3BrY clusters grow by conserving a markedly elongated shape, thus developing into precursors of the needle-like morphologies observed in experiments.

\section{Summary and Conclusions}

In our work, we have observed that 3BrY forms clusters, which are not completely amorphous but rather composed of ordered and disordered domains. On average, the most probable aggregates possess about 30\% of the molecules in a crystal-like configuration, arranged in a single domain. This means that the nucleation of such complex systems may proceed via the aggregation of small, already ordered building blocks. In this case, we find that the two-step nucleation process does not occur via the formation of an 
amorphous droplet, but rather by the formation of small, locally ordered, crystalline building 
blocks, which assemble into larger clusters.

To probe how these locally ordered 3BrY clusters can assemble into a larger nucleus, we have 
investigated the growth of a supercritical 3BrY aggregate from a supersaturated methanol solution.
The resulting cluster maintains a dynamical character, being able to fluctuate in size and shape and to exchange both monomers and small oligomers with the solution continuously.  
Remarkably, the growing cluster maintains a substantial fraction of disordered domains, and it is constantly involved in dynamical exchanges of molecules with the surrounding environment. Furthermore, we observe that the growth process produces an elongated structure, which clearly shows an aspect ratio typical of needle-like crystals observed in experiments \cite{harano2012heterogeneous}. We also want to point out that, within the range of supersaturation considered in this work we have not identified an alternative crystalline structure for 3BrY.

To conclude we note that the nucleation process uncovered in our work, instead of proceeding through a paradigmatic two-step mechanism in which a crystal is formed within the liquid-like precursor, involves a dynamical exchange of monomers and oligomers with the solution. Such assembly of partly ordered functional structures is responsible for the growth of a crystal-like phase. This process displays remarkable similarities with the self-assembly process of non-covalent supramolecular fibers in solution\cite{Albertazzi2014}. 
 
Our work provides details relevant for improving our current understanding of the crystallization of large organic molecules interacting through apolar stacking, such as 3BrY. Developing such an understanding is key to devise rational strategies to direct the assembly of organic materials characterized by specific structural and functional properties.   

\section{Acknowledgements}
The authors acknowledge the ETH Zurich Brutus cluster for the computational resources. M.P. acknowledges the VARMET European Union Grant ERC-2014-ADG-670227 and the National Centres of Competence in Research "Materials Revolution: Computational Design and Discovery of Novel Materials (MARVEL)" for funding.

\bibliography{3BrY}
\bibliographystyle{unsrt}

%
%

\end{document}